\documentclass[pra,showpacs,twocolumn,nofootinbib]{revtex4}

\usepackage{amsmath}
\usepackage{amsfonts}
\usepackage{amssymb}
\usepackage{epsfig}
\usepackage{dcolumn}
\usepackage{bm}

\newcommand{\norm}[1]{\Vert #1 \Vert}

\newcommand{\ket}[1]{\left\vert#1\right\rangle}

\newenvironment{proof}[1][Proof]{\begin{trivlist}
\item[\hskip \labelsep {\bfseries #1}]}{\end{trivlist}}
\newtheorem{theorem}{Theorem}
\newtheorem{corollary}[theorem]{Corollary}
\newtheorem{lemma}[theorem]{Lemma}

\begin{document}

\title{Performance of Deterministic Dynamical Decoupling Schemes:
Concatenated and Periodic Pulse Sequences}
\author{Kaveh Khodjasteh$^{1}$ and Daniel A. Lidar$^{2}$}
\affiliation{$^{1}$Department of Physics, University of Southern California, Los Angeles,
CA 90089}
\affiliation{$^{2}$Departments of Chemistry, Electrical Engineering, and Physics,
University of Southern California, Los Angeles, CA 90089}

\begin{abstract}
Dynamical decoupling can be used to preserve arbitrary quantum states
despite undesired interactions with the environment, using control
Hamiltonians affecting the system only. We present a system-independent
analysis of dynamical decoupling based on leading order decoupling error
estimates, valid for bounded-strength environments. Using as a key tool a
renormalization transformation of the effective system-bath coupling
Hamiltonian, we delineate the reliability domain of dynamical decoupling
used for quantum state preservation, in a general setting for a single
qubit. We specifically analyze and compare two deterministic dynamical
decoupling schemes -- periodic and concatenated -- and distinguish between
two limiting cases of fast versus slow environments. We prove that
concatenated decoupling outperforms periodic decoupling over a wide range of
parameters. These results are obtained for both ``ideal'' (zero-width) and
realistic (finite-width) pulses This work extends and generalizes our
earlier work, Phys. Rev. Lett. \textbf{95}, 180501 (2005).
\end{abstract}

\maketitle

\section{Introduction}

Arbitrary quantum state preservation is a fundamental imperative for quantum
information processing, but undesired interactions of a candidate quantum
system with uncontrollable external systems (the environment/bath) results
in poor control and fidelity loss. Even if the structure of these
interactions is approximately known, the statistical uncertainty in the
state of the environment invariably results in decoherence \cite{Zurek}.
While undesired couplings to the environment are inevitable (even at zero
temperature), strong control fields applied to the system can be used to
effectively manipulate the couplings. Nuclear magnetic resonance is an
excellent example where techniques such as refocusing and composite pulses
are readily used to generate reliable and high precision quantum dynamics
\cite{Freeman:book,vandersypen:04}. Similar in execution but applicable in a
generic setting, \emph{dynamical decoupling} (DD) is a method for the
effective renormalization of the system-bath interaction Hamiltonian via the
application of strong system-control fields. Usually the goal is the
cancellation of all coupling terms. In the context of quantum information
processing, DD can be used for feedback-free quantum error suppression
without encoding overhead \cite{Viola:99}.

Dynamical decoupling is most efficient against \emph{bounded environments},
when the pulse switching times are short on a scale set by the bath spectral
density high-frequency cutoff \cite{Viola:98,Facchi:05}, or when the
spectral density is rapidly decaying \cite{ShiokawaLidar:02}. Within these
assumptions different flavors of DD can be designed. Of course, technology
limits how strongly, rapidly and accurately we can modulate the system
Hamiltonian, and cool the system. Dynamical decoupling can be implemented
with the pulse sequence chosen deterministically, e.g., periodically \cite%
{Viola:98,ByrdLidar:01,pryadko:085321}, or randomly \cite{Facchi:03,Viola:05}%
. Randomized decoupling is expected to perform better in the case of
varying/fluctuating (effective) Hamiltonians while deterministic methods
perform better in cases where the undesired terms in the system-bath Hamiltonian
are sufficiently weak \cite{Santos:06,Viola:06}. Hybrid schemes with optimized
performance have also been considered \cite{Viola:05,kern:250501}. The analysis of
DD\ schemes is often performed within an interaction picture. Here we
consider explicitly the internal dynamics of the bath in terms of its effect
on the performance of DD.

Dynamical decoupling strategies, some of which can be derived from group
theoretical considerations \cite{Zanardi:98b}, are typically based on a
\emph{universal} DD pulse sequence \cite{Viola:99}: a short sequence of
unitary operators designed to completely cancel errors up to the first order
in the Magnus expansion \cite{Magnus:54}. Here we consider two deterministic
decoupling schemes: (i) In periodic DD (PDD), the universal decoupling
sequence is repeated periodically for the duration of the quantum state
preservation. (ii) In concatenated dynamical decoupling (CDD) \cite%
{KhodjastehLidar:04}, the universal pulse sequence is recursively embedded
within itself. We provide an analytic leading-order study of the performance
of the above strategies. Our first basic finding is a verification in
the DD-setting of
a result familiar from NMR, that even when ideal
(zero-width) pulses are used for decoupling, the corrections from second and
higher order Magnus terms impose an upper performance bound on PDD.

A central result of our approach is that the coupling terms responsible for
errors undergo an effective renormalization transformation by the externally
applied pulse sequences. This process is conveniently described via the
Magnus expansions for derivation of effective coupling Hamiltonians [see
Eqs.~(\ref{eq:b0})-(\ref{eq:bz}) below]. The renormalization approach leads
to the view of DD as a dynamical map, whose convergence to a fixed point
(ideally, the cancellation of the system-bath interaction) depends on
whether the norm of the coupling terms decreases under repeated applications
of the pulse sequence. In support of our earlier study \cite%
{KhodjastehLidar:04}, within the technological constraints of finite pulse
numbers and the bounds imposed by the applicability of DD in general, we
analytically prove the asymptotic superiority of CDD over PDD. In addition,
we present here new pulse sequences, inspired by the Trotter-Suzuki
expansion \cite{Suzuki:76}, with even better convergence properties than
CDD. In a more abstract setting, we show that in fact any application of
unitary operators on the system cannot cause an increase in the strength of
the undesired couplings.

Our conclusions are valid within the convergence domain of our expansions.
We find that the Magnus expansion itself sets the most stringent limit on
convergence domains, in the sense that it includes or coincides with the
regime of applicability of DD. In the worst case, this corresponds to the
limit of slow internal bath dynamics.

We also analyze robustness with respect to systematic pulse errors. The
decoupling error of any deterministic scheme is thus a result of the
environmental coupling errors, and errors in the decoupling operations. In
the case of realistic, imperfect pulses, we show that the performance of DD,
when pulses of finite width (or uniform error rate) are used, is determined
by a condition based on both the pulse switching times and pulse widths. In
this case, unless pulse profiles and timings are adjusted in case of
systematic pulse errors, even the first order terms (and thus dominating) in
the Magnus expansion will be non-zero.

While undesired coupling terms are responsible for fidelity loss, the
relationship between the two is complicated and depends on the details of
the environment and possible physical energy cutoffs. This motivates us to
perform a generic analysis based on operator-norm estimates. Our
conservative estimates provide a worst-case analysis for decoupling
performance. We expect these estimates to be useful as guidelines for
choosing and combining dynamical decoupling strategies when constraints such
as pulse switching times and pulse errors are considered.

This paper is organized as follows: in Section~\ref{UDD} we review the basic
universal DD cycle which suffices to decouple a qubit from an arbitrary
non-Markovian environment to first order in the Magnus expansion. In Section~%
\ref{UDD-analysis} we provide a detailed analysis of this sequence in terms
of the Magnus expansion, for both ideal (zero-width) and non-ideal
(finite-width) pulses. In Section~\ref{strategies} we compare two
deterministic decoupling strategies founded on the basic universal DD
sequence: periodic and concatenated sequences. We calculate a fidelity
measure associated with the two strategies and show that the concatenated
one strictly outperforms the periodic one. In Section~\ref{TSD} we introduce
a new decoupling strategy, based on the Trotter-Suzuki expansion. Even
though this decoupling sequence has implementations problems and is not
as robust as CDD, we find it interesting in light of its superior
convergence properties. This Section also includes a table which compares
the three deterministic decoupling strategies (PDD, CDD, and
Trotter-Suzuki), and clearly illustrates and summarizes their relative
performance. Finally, in Section~\ref{Thompson} we present a general result
concerning the behavior of error norms under pulse sequences: we show that
for sufficiently narrow pulses, pulse sequences cannot increase error norms.
This result has impact also on fault tolerance theory using quantum error
correcting codes. A summary and discussion is presented in Section~\ref%
{summary}. Supporting calculations can be found in the Appendices.

\section{Universal Dynamical Decoupling for a Qubit}

\label{UDD}

In the absence of driving terms, the terms in the system-bath interaction
Hamiltonian are responsible for decoherence and loss of quantum information.
Removal of these terms is sufficient (but not necessary \cite{LidarWhaley:03}%
) for preservation of arbitrary quantum states. Dynamical decoupling schemes
use strong and fast control Hamiltonians acting on the quantum system only,
to effectively remove/modify various terms in the system-bath interaction
Hamiltonian \cite{CDD-PRA:note1}. In particular, a pulse sequence designed to
remove every term in the interaction Hamiltonian is referred to as \emph{%
universal dynamical decoupling}. In this work, we focus on the universal DD
of a single qubit. Extensions to multiple qubits \cite{Viola:99} and higher
dimensional quantum systems exist \cite{WuByrdLidar:02,ByrdLidarWuZanardi:05}, but we will not consider these here. We use $X$, $Y$, and $Z$ to denote
the standard $2\times 2$ Pauli matrices
\begin{equation*}
\sigma_{x}=%
\begin{pmatrix}
0 & 1 \\
1 & 0%
\end{pmatrix}%
, \, \sigma_{y}=%
\begin{pmatrix}
0 & -i \\
i & 0%
\end{pmatrix}%
,\, \sigma_{z}=%
\begin{pmatrix}
1 & 0 \\
0 & -1%
\end{pmatrix}%
\end{equation*}
acting on a single qubit, and work in units of $\hbar =1$. System and
environment are assumed to inhabit different Hilbert spaces i.e., we do not
consider leakage, which can also be treated using DD\ \cite%
{WuByrdLidar:02,ByrdLidarWuZanardi:05}), and all Hamiltonian operators are
taken to be traceless without loss of generality.

Consider a qubit with a Hamiltonian
\begin{equation}
H(t)=H_{\text{ctrl}}(t)+H_{e}(t),
\end{equation}%
where $H_{\text{ctrl}}$ refers to a time-dependent controllable system-only
part and $H_{e}$ includes all other terms, i.e., the internal bath, internal
system, and interaction Hamiltonians:%
\begin{equation}
H_{e}=H_{B}\otimes I_{S}+I_{B}\otimes H_{S}+H_{SB}.  \label{eq:H_e}
\end{equation}%
Here $I$ denotes the identity operator. We have implicitly excluded a
pure-bath Hamiltonian term $H_{B}^{0}\otimes I_{S}$ satisfying $%
[H_{B}^{0}\otimes I_{S},F]=0$, where $F$ is any element of the Lie algebra
generated by $H_{SB}$, $H_{B}\otimes I_{S}$, and $I_{B}\otimes H_{S}$. The
reason is that such a term on the one hand does not impact the system
dynamics, but on the other hand will increase the operator norm that arises
below in our calculations of decoupling errors. With this in mind, ideally
one would like to have $H_{e}=0$. The \textquotedblleft error
Hamiltonian\textquotedblright\ $H_{e}$ can always be expanded as:
\begin{equation}
H_{e}=B_{0}\otimes I_{S}+B_{X}\otimes X+B_{Y}\otimes Y+B_{Z}\otimes Z
\end{equation}%
where $B_{\alpha }$ ($\alpha =0,X,Y,Z$) are operators acting on the
environment. We are allowing the $B_{\alpha }$ to include the identity
operator, i.e., \emph{from now on we are incorporating }$H_{S}$ \emph{into} $%
H_{SB}$. I.e., assuming
\begin{equation}
H_{S}=\sum_{\alpha =x,y,z}\omega _{\alpha }\sigma _{\alpha },
\end{equation}%
with $\omega _{\alpha }$ all non-zero frequencies, and writing
\begin{equation}
H_{SB}=\sum_{\alpha =x,y,z}b_{\alpha }\otimes \sigma _{\alpha },
\end{equation}%
yields
\begin{eqnarray}
B_{\alpha } &=&\omega _{\alpha }I_{B}+b_{\alpha }\quad \alpha \in \{x,y,z\},
\notag \\
B_{0} &=&I_{B}+H_{B}.
\end{eqnarray}%
Note that the first term in $H_{e}$, $B_{0}\otimes I_{S}$, is a
pure-environment term and simply generates the environment's internal
dynamics. It also includes the global phase generating term $I_{B}\otimes
I_{S}$. Obviously if $b_{X}=b_{Y}=b_{Z}=0$, then $H_{e}=B_{0}\otimes I_{S}$
has no effect on the system dynamics.

Universal DD of a qubit with ideal pulses removes every term in $H_{e}$
except $B_{0}\otimes I_{S}$, by applying the following pulse sequence: {%
\texttt{f}$X$\texttt{f}$Z$\texttt{f}$X$\texttt{f}$Z$}, where $\mathtt{f}$
denotes a \textquotedblleft pulse-free\textquotedblright\ period of fixed
duration. The pulses are generated by $H_{\text{ctrl}}$. This universal DD
sequence is a simplification of $(I_{S}\mathtt{f}I_{S})(X\mathtt{f}X)(Y%
\mathtt{f}Y)(Z\mathtt{f}Z)$, where $X,Y,Z$ and the identity $I_{S}$
represent the \emph{decoupling group} ${\mathcal{G}}$ on the qubit. The
universal decoupling group has the property that for every Hamiltonian $H$
acting on the system, the sum $\sum_{\{P_{i}\in {\mathcal{G}}%
\}}P_{i}HP_{i}^{\dag }$ acts trivially on the system \cite{Viola:99}. Since
this sum is the leading order generator in the Magnus expansion, the
universal DD sequence completely removes any system-bath interaction to
first order (we revisit this in detail in subsection \ref{magnus}).

A complete analysis of the performance of DD needs to take into account
details of the environment (participating modes, energy cutoffs,
temperature, (dis)equilibrium, etc.) but we shall minimize these
considerations and focus on the model-independent features of DD. Our
analysis is mathematically constrained by convergence domains that are
explicit in our approximations. We expect certain unbounded systems such as
bosonic environments to be within the realm of our theoretical framework
after the introduction of spectral cutoffs.

\section{Analysis of the Universal Decoupling Pulse Sequence: The
Renormalization Transformation}

\label{UDD-analysis}

In this section we derive the transformation of various Hamiltonian terms
under the basic universal DD pulse sequence. We first consider ideal pulses,
then amend our discussion to allow for non-ideal (finite width) pulses. We
will see that the system-bath interaction Hamiltonian is effectively
renormalized under the DD pulse sequence. As long as this renormalization
transformation is norm-reducing, the DD procedure is effective.

\subsection{Ideal Pulses}

Let us first assume that the pulses used are ideal, i.e., infinitely strong
and narrow. For example, $H_{\text{ctrl}}(t)=\frac{\pi}{2}\delta(t-t_{0})X$
generates an ideal $X$ pulse at time $t_{0}$ ($\delta$ is the Dirac $\delta$%
-function). The propagator corresponding to free evolution $\mathtt{f} $ of period $\tau_{0}$ is
given by:
\begin{equation}
U_{\mathtt{f}}=\exp(-i\tau_{0}H^{(0)}),\quad H^{(0)}:=H_{e}.
\end{equation}
To obtain the cycle propagator we use the identity $%
Me^{A}M^{-1}=e^{MAM^{-1}} $, valid for any operator $A$ and invertible $M$.
Define
\begin{align}
H_{1} & \equiv B_{0}\otimes I+B_{X}\otimes X+B_{Y}\otimes Y+B_{Z}\otimes
Z=IH_{e}I,  \notag \\
H_{2} & \equiv B_{0}\otimes I+B_{X}\otimes X-B_{Y}\otimes Y-B_{Z}\otimes
Z=XH_{e}X,  \notag \\
H_{3} & \equiv B_{0}\otimes I-B_{X}\otimes X+B_{Y}\otimes Y-B_{Z}\otimes
Z=YH_{e}Y,  \notag \\
H_{4} & \equiv B_{0}\otimes I-B_{X}\otimes X-B_{Y}\otimes Y+B_{Z}\otimes
Z=ZH_{e}Z,  \label{eq:Hi}
\end{align}
The free evolution propagator can then also be written as {\small \texttt{f}}%
$=\exp(-i\tau_{0}H_{1})$. Using Eqs.~(\ref{eq:Hi}) we can write the total
propagator corresponding to the universal DD cycle, $(I\mathtt{f}I)(X\mathtt{%
f}X)(Y\mathtt{f}Y)(Z\mathtt{f}Z)$, in terms of four effective Hamiltonians
describing the different evolution segments, as:
\begin{equation}
U_{1}=e^{-i\tau_{0}H_{1}}e^{-i\tau_{0}H_{2}}e^{-i\tau_{0}H_{3}}e^{-i\tau
_{0}H_{4}}.  \label{ali:expo}
\end{equation}
A time-varying piecewise constant Hamiltonian, $H(t)$ varying over four
intervals each of length $\tau_{0}$, can generate this propagator. The total
propagator can then be used to define the effective Hamiltonian $H^{(1)}$:
\begin{equation}
U_{1}=\prod_{i=1}^{4}\exp(-i\tau_{0}H_{i}^{(0)})=:\exp(-i4\tau_{0}H^{(1)}),
\label{eq:U1}
\end{equation}
where we have added superscripts $(0)$ to the Hamiltonians $H_{i}$ of Eq.~(%
\ref{eq:Hi}), in anticipation of the concatenation procedure that we
consider in subsection~\ref{CDD}.

\subsection{Magnus Expansion}

\label{magnus}

The cycle propagator $U_{1}$ [Eq.~(\ref{eq:U1})] can be approximated using a
Magnus expansion (for an alternative method of analysis that is particularly
useful for the design of periodic sequences of soft pulses see Ref.~\cite%
{sengupta:037202}). Consider a time-dependent Hamiltonian $H(t)$ generating
the propagator $U(t)$ from time $0$ to $t$. In the Magnus expansion (a type
of cumulant expansion) we have
\begin{equation}
U(t)=\exp \sum_{i=1}^{\infty }A_{i}(t),
\end{equation}%
with $A_{1}$ and $A_{2}$ given by:
\begin{align}
A_{1}& =-i\int_{0}^{t}dt_{1}H(t_{1}),  \label{eq:A1-mag} \\
A_{2}& =-\frac{1}{2}\int_{0}^{t}dt_{1}%
\int_{0}^{t_{1}}dt_{2}[H(t_{1}),H(t_{2})].  \label{eq:A2-mag}
\end{align}%
Higher order terms are given by higher order commutator expressions \cite%
{Magnus:54}. A recent bound for the convergence radius of the Magnus
expansion \cite{Moan:01} translates in our case into $\max {\Vert H(t)\Vert }%
t<2$.\footnote{%
Throughout this work we use $\left\Vert A\right\Vert $ to denote a unitary
invariant operator norm, e.g., the maximum eigenvalue for traceless
operators, or the absolute difference between the smallest and largest
eigenvalues \cite{Bhatia:book}.} In many situations $\Vert B_{0}\Vert $
(norm of the environment's internal Hamiltonian) is expected to dominate the
Hamiltonian and we may as well use $\Vert B_{0}\Vert t\lesssim 1$ as a
conservative convergence domain.
This bound can, however, be superficial since not all degrees of freedom of
the environment might actually be involved in the dynamics. For example, in
a spin bath, bath spins far away from the system will not immediately
contribute to the dynamics but this will nonetheless increase $\Vert
B_{0}\Vert $ without changing the real convergence radius. A precise
analysis of the actual convergence radius requires us to estimate the
next-to-leading-order terms -- see Appendix~\ref{appA}.

For the piecewise constant evolution of the Hamiltonian in Eq. (\ref%
{ali:expo}), we can calculate $A_{1}$ and $A_{2}$ in terms of $H(t)=$ $%
\{H_{j}$ for $(j-1)\tau _{0}\leq t\leq j\tau _{0}\}_{j=1}^{4}$, i.e., the
sign-flipped Hamiltonians appearing in Eq.~(\ref{ali:expo}):
\begin{align}
A_{1}^{(1)}& =-i\tau _{0}(H_{1}^{(0)}+H_{2}^{(0)}+H_{3}^{(0)}+H_{4}^{(0)})
\label{eq:a1} \\
A_{2}^{(1)}& =-\frac{1}{2}\tau
_{0}^{2}\sum_{1=i<j=4}[H_{i}^{(0)},H_{j}^{(0)}].  \label{eq:a2}
\end{align}%
Again, the superscripts are included in anticipation of the CDD analysis
below. Note that the pure-environment parts of $A_{i}^{(1)}$, i.e., terms of
the form $B\otimes I$, have no effect on the dynamics of the qubit to first
order in $\tau _{0}$, but do have an effect to second order in $\tau _{0}$,
through the commutator terms. Clearly, pure-environment terms are not
renormalized under the DD procedure.

Using Eqs.~(\ref{eq:Hi}),(\ref{eq:a1}),(\ref{eq:a2}) we find:
\begin{eqnarray}
A_{1}^{(1)}& =&-i(4\tau _{0})B_{0}\otimes I  \notag \\
A_{2}^{(1)}& =&4\tau_0^2[B_0,B_X]\otimes X  \notag \\
&& + 2\tau_0^2\left([B_0,B_x]-i\{B_X,B_Z\}\right)\otimes Y  \label{eq:magnus}
\end{eqnarray}
\emph{This shows that while to first order in the Magnus expansion (the }$%
A_{1}^{(1)}$\emph{\ term) the universal decoupling cycle completely removes
the coupling to the environment, there is a leading second order correction
due to }$A_{2}^{(1)}$\emph{\ in which the coupling to the environment has
not been removed.}

Note that a pure-environment term appears only in $A_{1}^{(1)}$ and, due to
our particular choice of DD sequence, {\small \texttt{f}$X$\texttt{f}$Z$%
\texttt{f}$X$\texttt{f}$Z$}, there is no term involving $\otimes Z$ in $%
A_{2}^{(1)}$.

We now define two norms which will play a central role in our analysis:
\begin{align}
\beta & :=\left\Vert B_{0}\right\Vert <\infty \\
J& :=\max (\left\Vert B_{X}\right\Vert ,\left\Vert B_{Y}\right\Vert
,\left\Vert B_{Z}\right\Vert )<\infty .  \label{eq:Jb}
\end{align}%
Recall that $B_{\alpha }=\omega _{\alpha }I_{B}+b_{\alpha }$ for $\alpha \in
\{x,y,z\}$, and $B_{0}=I_{B}+H_{B}$; unless otherwise specified, we assume that $%
J<\beta $ in order to simplify our convergence arguments. This conservative assumption is
reasonable for systems where only a small number of environment particles (or degrees of freedom)
are coupled to a given qubit
(such as electron spins coupled to a nuclear spin bath \cite%
{khaetskii:186802}) -- which translates into a small $\Vert b_{\alpha }\Vert
$ -- whereas no restriction exists on the environment self-Hamiltonian (%
\emph{e.g.}, on the number of particles). The distinction between $J$ and $%
\beta $ is a reflection of the different roles $H_{SB}$ and $H_{B}$ play in
the dynamics of the system.
In simple terms, $J$ quantifies the direct coupling strength while $\beta $
quantifies typical bath frequencies. Consider, e.g., the simple case of
a spin qubit coupled to another spin-$1/2$ particle via a Heisenberg coupling: $H_{e}=\omega
Z_{B}\otimes I+c(X_{B}\otimes X+Y_{B}\otimes Y+Z_{B}\otimes Z)$, where $c$ is
the coupling coefficient. In this case we have: $J=O(c)$ and $\beta
=O(\omega )$.

The Magnus terms can be bounded using these quantities: $\Vert A_{1}^{(1)}
\Vert=O(\tau _{0}\beta )$ and $\Vert A_{2}^{(1)} \Vert=O(\tau^2_{0}\beta J)$%
. Higher order Magnus terms, $A_{i>2}^{(1)}$, will contain all orders of $%
\tau _{0}^{i}J^{k}\beta ^{i-k}$ where $1\leq k\leq i-1$, and the leading
term is always given by $O(\tau _{0}^{i}J\beta ^{i-1})$ since $J<\beta $. As
long as $\tau _{0}\beta \ll 1$ we can safely neglect $A_{i>2}^{(1)}$:
\begin{equation}
||A_{i>2}^{(1)}||\ll ||A_{2}^{(1)}||<||A_{1}^{(1)}||\text{.}  \label{eq:A2i}
\end{equation}%
Note that our derivations are based on the separation of the coupling terms $%
B_{0}\otimes I$ and $B_{X}\otimes X+B_{Y}\otimes Y+B_{Z}\otimes Z$ in the
error Hamiltonian $H_{e}$. A similar separation can be done for decoupling
schemes on systems other than a single qubit.

The approximate effective Hamiltonian $H^{(1)}$ corresponding to the basic
dynamical decoupling cycle, Eq.~(\ref{eq:U1}), is now:
\begin{equation}
H^{(1)}\approx \frac{1}{-i4\tau _{0}}(A_{1}^{(1)}+A_{2}^{(1)}):=\sum_{\alpha
}B_{\alpha }^{(1)}\otimes \sigma _{\alpha }.
\end{equation}%
The renormalized environment operators $B_{\alpha }^{(1)}$, which can easily
be read off from Eqs.~(\ref{eq:magnus}), are the main result of the DD
procedure. The important message emerging from the analysis in this
subsection is that even when ideal (infinitely strong and narrow) pulses are
used, the universal decoupling cycle only removes (the system-bath component
of) the lowest order Magnus term, and renormalizes the higher order Magnus
terms. The success of dynamical decoupling ultimately depends on whether the
mapping to the renormalized environment operators is significantly
norm-decreasing, an issue we address in detail below.

\subsection{Pulses of Finite Width}
\label{finite-width}

Ideal pulses that act as system-only unitary operators are simplified
mathematical abstractions. In this subsection we model and analyze the
effect of the finite width of pulses in decoupling. For simplicity we
consider rectangular pulses $P$ with a width $\delta $. The ideal
finite-width pulse is,
\begin{equation}
P=\exp (-i\delta H_{P}),
\end{equation}%
where $H_{P}$ is a fixed control Hamiltonian. For realistic pulses we must include
$H_{e}$ in the pulse propagator:
\begin{equation}
U_{P}=\exp (-i\delta (H_{P}+H_{e})).
\end{equation}%
We call such a pulse \textquotedblleft non-ideal\textquotedblright .
Extremely narrow pulses with $||H_{P}||\gg ||H_{e}||$ are thus desirable to
minimize the effect of the unwanted terms in the pulse Hamiltonian. Here we
build upon the approximation of instantaneous pulses in subsection~\ref%
{magnus}, by decomposing the non-ideal pulses into products of the ideal
unitary operator of the pulse $P$ and some effective pulse error unitary
operator $E_{P}$. We explicitly approximate the operators $E_{P}$ for
rectangular pulses on a qubit, but the decomposition of the actual pulse
into an ideal pulse and a \textquotedblleft pulse error\textquotedblright\
unitary can be reproduced for other pulse shapes as well.

The periods of the universal dynamical decoupling cycle need to be adjusted
in order to incorporate the time delays associated with finite pulse widths.
Therefore, assume that all free evolution periods, with propagator $U_{%
\mathtt{f}}$, are adjusted to length $\tau_0 -\delta $. The propagator for
the cycle can then be written as:
\begin{align}
U^{(1)}& =U_{\mathtt{f}}U_{X}U_{\mathtt{f}}U_{Z}U_{\mathtt{f}}U_{X}U_{%
\mathtt{f}}U_{Z}  \notag \\
& =U_{\mathtt{f}}E_{X}XU_{\mathtt{f}}XE_{Z}^{\prime }YU_{\mathtt{f}%
}YE_{X}^{\prime }ZU_{\mathtt{f}}ZE_{Z}  \notag \\
& =:U_{\mathtt{f}_{1}}E_{X}U_{\mathtt{f}_{2}}E_{Z}^{\prime }U_{\mathtt{f}%
_{3}}E_{X}^{\prime }U_{\mathtt{f}_{4}}E_{Z}  \label{eq:seq}
\end{align}%
where
\begin{equation}
U_{\alpha }=e^{-i\delta (\eta \sigma _{\alpha }+H_{e})}\qquad \alpha =X,Z,
\end{equation}%
$\delta \eta =\pi /2$, and, in order to fit the formulation of subsection~%
\ref{magnus}, we have defined the pulse-error operators as follows:
\begin{align}
E_{X}X:=U_{X},& \hspace{1cm}YE_{X}^{\prime }Z:=U_{X},  \label{eq:eX} \\
ZE_{Z}:=U_{Z},& \hspace{1cm}XE_{Z}^{\prime }Y:=U_{Z}.  \label{eq:eZ}
\end{align}%
Note that since the errors $E_{\alpha }$ are unitary and are produced during
an interval $\delta $, we may formally associate them with effective
Hamiltonians defined through
\begin{equation}
E_{\alpha }=:\exp (-i\delta H_{E,\alpha }).
\end{equation}%
Using these definitions, Eq.~(\ref{eq:seq}) is equivalent to the evolution
due to a piecewise constant Hamiltonian $H(t)$, given by the sequence $%
\{H_{1},H_{E,X},H_{2},H_{E,Z}^{\prime },H_{3},H_{E,X}^{\prime
},H_{4},H_{E,Z}\}$ with $H_{i}$ given in Eqs.~(\ref{eq:Hi}), at appropriate
times. We ignore terms of order $\delta ^{2}||B_{\alpha }||^{2}$ and $\delta
\tau _{0}\Vert B_{\alpha }\Vert ^{2}$, which allows us to treat the
components of $H_{E,\alpha }$ as c-numbers instead of operators, since no
commutators will be involved. Using Eqs.~(\ref{eq:A1-mag}),(\ref{eq:A2-mag}%
), we can repeat the calculation of $A_{1}^{(1)}$ and $A_{2}^{(1)}$, this
time including the pulse segments with $H_{E,\alpha }$ as their effective
Hamiltonians, and consider the limit of narrow pulses, $\delta \ll \tau _{0}$%
. In this limit, we can safely truncate the Magnus expansion after $%
A_{2}^{(1)}$, {provided we assume:}
\begin{equation}
c\tau _{0}\beta +d\frac{\delta }{\tau _{0}}\ll 1,  \label{eq:taunb'}
\end{equation}%
where $c$ and $d$ are numerical factors of $O(1)$ (recall that $\beta
:=\left\Vert B_{0}\right\Vert $). This inequality is derived in Appendix~\ref%
{appB}. Note that it implies an optimal pulse interval $\tau _{0}=\sqrt{%
d\delta /c\beta }$, which minimizes the left-hand side of (\ref{eq:taunb'})
for given $\beta $ and a fixed minimal pulse width $\delta $. Note further
that Ineq.~(\ref{eq:taunb'}) is not as strict as the condition for the
convergence of the Magnus expansion, which reads (when $\beta \gg J$): $%
\beta T <1$, where $T$ is the total experiment duration.

The components of the effective Hamiltonian can be calculated explicitly:
\begin{align}
B_{0}^{(1)}& =B_{0},  \label{eq:B0(n)-fw} \\
B_{X}^{(1)}& =i(\tau _{0}-\delta )[B_{0},B_{X}^{(0)}]+\frac{\delta }{ \tau
_{0}}(\frac{1}{2}B_{X}^{(0)}-\frac{1}{\pi }B_{Y}^{(0)}), \\
B_{Y}^{(1)}& =\frac{i}{2}\tau
_{0}([B_{0},B_{Y}^{(0)}]-i\{B_{X}^{(0)},B_{Z}^{(0)}\})  \notag \\
& +\frac{i}{2}\delta ([B_{0},B_{Y}^{(0)}]-2i\{B_{X}^{(0)},B_{Z}^{(0)}
\}-2iB_{Z}^{(0)}B_{X}^{(0)})  \notag \\
&+\frac{1}{\pi }\frac{\delta }{\tau _{0}} B_{Z}^{(0)}, \\
B_{Z}^{(1)}& =i\delta \left[ \frac{2}{\pi }
B_{X}^{(0)}(B_{Z}^{(0)}+B_{X}^{(0)})+B_{Y}^{(0)}B_{X}^{(0)}\right] + \frac{%
\delta }{\tau _{0}}B_{Z}^{(0)}  \label{eq:BZ(n)-fw}
\end{align}
The only modifications associated with the pulse width $\delta $ are of
order $O(J(\delta J+\delta/\tau _{0}))$, associated with the new small
parameters $\delta J$ and $\delta /\tau _{0}$. \emph{In this case the
decoupling is not exact and even the first order Magnus terms contribute
decoupling errors of order} $\delta J$, the ``per-pulse-error''. This is an important effect which
will adversely affect decoupling schemes not designed to compensate for such
finite pulse-width errors. 

We note that it is possible to design a piece-wise constant profile for the
control Hamiltonian $H_{\text{ctrl}}$ such that the first order Magnus
corrections due to systematic errors in the control Hamiltonian are zero
\cite{Viola:02}. The separation of non-ideal pulses into ideal and error
pulses, as above, still applies to this \textquotedblleft Eulerian
decoupling\textquotedblright\ scheme, as do most bounds we obtain here.
Finally, we note that treatments of decoupling and refocusing errors similar
to the above have been pursued in an NMR-specific setting \cite{Haeberlen:book}.

\section{Decoupling Strategies}\label{strategies}

The universal DD cycle results in segments of evolution with Hamiltonians $%
H_{i}$ such that $\sum H_{i}$ acts trivially on the system. The derivations
of the previous section show how the actual overall propagator contains
higher order corrections that \emph{do not} act trivially on the qubit.
Nonetheless, the basic universal DD cycle provides us with the building
blocks of general decoupling schemes that optimize decoupling performance
with respect to constraints in switching times and pulse precision.
Numerical simulations comparing and discussing some of the schemes
(deterministic, randomized, or hybrid) are available
\cite{Santos:05,Viola:06,kern:250501} (ideal pulses) and
\cite{Santos:06} (ideal and non-ideal pulses), and in this section we focus on
system-independent analytic arguments. We deviate from our abstract
treatment of bath operators by considering two limiting cases for the
coupling strengths of system-bath and pure-bath, namely: (i) $J<\beta $ and
(ii) $\beta \ll J$ [recall Eq.~(\ref{eq:Jb})]. In case (i), the coupling to
the environment induces slow dynamics while the environment itself has fast
dynamics. In case (ii), the coupling to the environment is dominant
but relatively stable due to the environment's slow internal
dynamics. This regularity makes case (ii) more attractive for
dynamical decoupling, or similar methods \cite{Yao:06}, while case (i)
is a worst case scenario. As will be noted however the presence of
higher-order commutators blurs out the distinction between the two
cases in higher orders of concatenated decoupling (Subsection
\ref{CDD}). Nonetheless, both cases are still within the convergence
domain of Magnus expansion ($\Vert H_e \Vert T<1$, see
subsection~\ref{magnus}). Outside the convergence domain
(e.g. corresponding to a longer duration of the experiment),
deterministic decoupling might be replaced by randomized decoupling
methods, for which there is some evidence of better performance
\cite{Santos:06}. In practice the co-existence of various bath regimes
makes hybrid decoupling methods a practical choice for long-time
decoupling \cite{Santos:05}.

\subsection{Error Phase}\label{errorphase}

We require a measure of fidelity to quantify the performance of DD. To this
end we define the \emph{error phase} corresponding to a propagator $%
U=e^{-iTH_{e}}$, describing an evolution of total duration $T$ generated by
an effective Hamiltonian $H_{e}$, as
\begin{equation}
\Phi :=Th,
\end{equation}%
where
\begin{equation}
h=\Vert H_{e}-B_{0}\otimes I\Vert
\end{equation}%
is the norm of the non-pure-environment part of $H_{e}$, which is
effectively (like $J$) a measure of the coupling strength. The
pure-environment part is explicitly excluded from the error phase since it
does not affect the fidelity $f$ (state overlap between ideal and decoupled
evolution) up to the leading order in our expansion. In this manner the
error phase now simply connects the coupling terms in the Hamiltonian to the
infidelity and decoherence. Indeed, for small error phases the infidelity, $%
1-f$, depends monotonically on $\Phi $ -- see Eq.~(\ref{eq:fid}) below.

In any physical implementation of dynamical decoupling we are limited by
technological constraints. Let $N$ denote the number of pulses used during a
decoupling experiment of duration $T$. Normally this number is bounded above
due to a minimum pulse switching time $\tau _{\text{min}}$: $N<T/\tau _{%
\text{min}}$. A basic pulse width $\delta $ is used when required, to
characterize the systematic error due to non-zero pulse widths. These
technological constraints are incorporated below by evaluating the fidelity
gain due to decoupling in terms of $N$, $T$, and $\tau _{0}$.

\subsection{Periodic Decoupling}

In periodic DD for a qubit, a basic universal sequence, such as {\small
\texttt{f}$X$\texttt{f}$Z$ \texttt{f}$X$\texttt{f}$Z$}, is repeated
periodically over the whole interval $T$. If $N$ pulses are used, there are
then $N$ intervals of length $\tau _{0}=T/N$ that correspond to the free
evolution periods, and $N/4$ repetitions of the basic sequence. The
effective Hamiltonian for the total interval, is obtained from the results
of the previous section, as long as we are within the convergence limit of
the Magnus expansion, given by $T\Vert H_{e}\Vert <1$. Limiting the Magnus
expansion to the first two terms, $A_{1}$ and $A_{2} $, the propagator for
the total evolution is given by
\begin{align}
U& \approx \exp [A_{1}^{(1)}+A_{2}^{(1)}]=\exp (-iTH^{(1)})  \notag \\
& =\exp \left[ -iT\left( B_{0}\otimes I+i\tau _{0}\sum_{\alpha,\beta,\gamma
    \in {0,X,Y,Z}} F_{\alpha \beta
}^{\gamma }B_{\alpha }B_{\beta }\otimes \sigma _{\gamma }\right) \right]
\label{eq:UPDD}
\end{align}%
where the effective coupling coefficients $F_{\alpha \beta }^{\gamma }$ are given in
Eqs.~({\ref{eq:magnus})} and can be calculated for any decoupling
scheme (not just the universal sequence).

For brevity we introduce the parameter $G:=\max(J,\beta)$. We read off the error phase from Eq.~(\ref{eq:UPDD}) as
\begin{equation}
\Phi _{\mathrm{PDD}\text{(i)}}^{(1)} = T h = O(TJ\tau _{0} G ). \label{eq:PhiPDD}
\end{equation}%
The above estimate for ideal pulses will be modified to the following if
rectangular pulses of width $\delta $ are used:
\begin{equation}
\Phi _{\mathrm{PDD}\text{(i)}}^{(1)}(\delta )=O[TJ(\tau _{0}G +\delta /\tau _{0})].
\label{eq:widthpdd}
\end{equation}%
Our estimates show that in if ideal pulses are used, using a higher number of pulses, at fixed $T$,
leads to monotonic improvement in the error phase. I.e., the error phase is proportional to the
pulse interval $\tau _{0} $. Technology sets a lower limit $\tau _{\text{min}}$ on $\tau _{0}$,
which implies that the infidelity is bounded  from below by a monotonic function of $\tau
_{\text{min}}TJG$ in the ideal pulse limit, and the fidelity gain
scales with the number of pulses $N$. From
Eqs.~(\ref{eq:widthpdd}) for non-ideal pulses, we expect the optimal pulse interval to be given
approximately by $\tau _{0}=\max \{\tau _{\text{min}},(\delta /G )^{1/2}\}$. In these expressions
we have assumed that the pulse width $\delta $ is already at the technological lower limit.

\subsection{Concatenated Decoupling}

\label{CDD}

\subsubsection{Definition}

Significant improvement over periodic DD can be obtained by constructing a
concatenated sequence, i.e., by recursively embedding the basic universal DD
cycle within itself \cite{KhodjastehLidar:04}. This is done in the following
manner:
\begin{align}
\mathtt{p}_{0}& =\mathtt{f}  \notag \\
\mathtt{p}_{1}& =\mathtt{p}_{0}X\mathtt{p}_{0}Z\mathtt{p}_{0}X\mathtt{p}_{0}Z
\notag \\
& \vdots  \notag \\
\mathtt{p}_{n}& =\mathtt{p}_{n-1}X\mathtt{p}_{n-1}Z\mathtt{p}_{n-1}X\mathtt{p%
}_{n-1}Z.
\end{align}%
Here $\mathtt{p}_{0}$ (no pulses) is of duration $\tau _{0}$, $\mathtt{p}%
_{1} $ is of duration $\tau _{1}=4\tau _{0}$ (in the limit of ideal,
zero-width pulses), and $\mathtt{p}_{n}$ is of duration $\tau _{n}=4\tau
_{n-1}=4^{n}\tau _{0}$ ($n$ levels of concatenation).

As we are about to show, PDD dramatically outperforms CDD over a wide
parameter range. At an intuitive level, this is attributable to the fact
that CDD, with its self-similar structure, has error correcting capabilities
at multiple resolution levels, whereas PDD allows errors to accumulate
essentially as a random walk.

First, however, let us remark that the aperiodic sequence $\mathtt{p}_n$ may
be simplified using Pauli matrix product identities when Pauli operators
appear in succession ($XY=Z$ and cyclic permutations), in the same manner
that the universal decoupling sequence is simplified from $(I_{S}\mathtt{f}%
I_{S})(X\mathtt{f}X)(Y\mathtt{f}Y)(Z\mathtt{f}Z)$ to $\mathtt{p}_{1}$. The
reduction in the number of pulses gained by such algebraic cancellations
might not be strictly advantageous in a practical setting, since
experimentally it is not always possible to generate rotations around all
three axes. Moreover, the simplification does not change the asymptotic
behavior of the number of pulses as a function of the concatenation level $n$%
. The simplification does affect the formal recursive structure of the
sequence. This has no physical effect when decoupling pulses are ideal. But
with realistic, finite width pulses, this loss of self-similarity might
adversely affect the robustness of the pulse sequence against systematic
errors. On the other hand one could argue that the product of two slightly
wrong pulses is worse than one, which would be an argument in favor of
simplification. A clear decision one way or the other must be made in a
context-specific setting.

\subsubsection{Effective Hamiltonian}

Due to its recursive definition, the propagator corresponding to $p_{n-1}$
is generated by an effective Hamiltonian $H_{e}^{(n-1)}$, which is then
decoupled with $p_{n}$ and in turn generates $H_{e}^{(n)}$. The interval
length is multiplied by $4$ in each such recursive step in which the
effective Hamiltonian is renormalized. The truncation of the Magnus
expansion beyond the first two terms must be justified, so that the higher
order terms do not accumulate as the concatenation level goes up -- we do
this in Appendix~\ref{appA}. We can then construct the higher order
effective Hamiltonians by truncating the Magnus expansion and recursively
obtaining $A_{i}^{(n)}$ from $H_{e}^{(n-1)}$. In the following, $H_{i}^{(n)}$
are constructed as in Eqs.~(\ref{eq:Hi}) and are reproduced each time from
the Magnus expansion:
\begin{eqnarray}
H_{e}^{(n)} &=&\frac{i}{\tau _{n}}(A_{1}^{(n)}+A_{2}^{(n)})  \notag \\
&=&\frac{1}{4}(H_{1}^{(n-1)}+H_{2}^{(n-1)}+H_{3}^{(n-1)}+H_{4}^{(n-1)})
\notag \\
&&-\frac{i}{32}\tau _{n}\sum_{1=i<j=4}[H_{i}^{(n-1)},H_{j}^{(n-1)}],
\label{eq:hrec}
\end{eqnarray}%
The sum of the operators $H_{i}^{(n-1)}$ is independent of $\tau _{0}$ and
contributes to the pure-environment part. Nonetheless, $H_{e}^{(n)}$
contains the commutator terms that do include $4^{n}\tau _{0}$ (and
contribute to the error-terms acting on the system). The commutator $%
[H_{i}^{(n-1)},H_{j}^{(n-1)}]$ must compensate for this exponential growth.

For the qubit case we can derive the explicit form of the sequence of
effective Hamiltonians $H_{e}^{(n)}$, by finding the ${B}_{\alpha }^{(n)}$.
We already have the first step of the concatenation in Eq.~(\ref{eq:magnus}%
). This also serves to initialize the recursion. Proceeding recursively we
obtain, for the next iteration:
\begin{eqnarray}
B_{0}^{(1)} &\mapsto &B_{0}^{(2)}=B_{0},  \notag \\
B_{X}^{(1)} &\mapsto &B_{X}^{(2)}=i\tau _{1}([B_{0}^{(1)},B_{X}^{(1)}] =
i\tau _{1}[B_{0},B_{X}^{(1)}],  \notag \\
B_{Y}^{(1)} &\mapsto &B_{Y}^{(2)}=i\tau _{1}\frac{1}{2}%
([B_{0}^{(1)},B_{Y}^{(1)}]-i\{B_{X}^{(1)},B_{Z}^{(1)}\})  \notag \\
&&= i\tau _{1}\frac{1}{2}[B_{0},B_{Y}^{(1)}],  \notag \\
B_{Z}^{(1)} &\mapsto &B_{Z}^{(2)}=0.
\end{eqnarray}%
Therefore, for general $n$:
\begin{align}
B_{0}^{(n)}&=B_{0}, & \hspace{1cm}\text{for }n\geq 0  \label{eq:b0} \\
B_{X}^{(n)}&=(i\tau _{n-1})[B_{0},B_{X}^{(n-1)}], & \hspace{1cm}\text{for }%
n\geq \text{$1$}  \label{eq:bx} \\
B_{Y}^{(n)}&=\frac{1}{2}(i\tau _{n-1})[B_{0},B_{Y}^{(n-1)}], & \hspace{1cm}%
\text{for }n\geq \text{$2$}  \label{eq:by} \\
B_{Z}^{(n)}&=0. & \hspace{1cm}\text{for }n\geq \text{$1$}  \label{eq:bz}
\end{align}
These equations capture the essence of the \emph{renormalization
transformation} the error terms experience under the pulse sequence.

\subsubsection{Convergence and Performance}

Next we study the convergence conditions of the latter recursive relations.
Let us define
\begin{equation}
h^{(n)}:=\max \{\Vert B_{X}^{(n)}\Vert ,\Vert B_{Y}^{(n)}\Vert \}.
\end{equation}%
Since $H_{e}^{(n)}=\sum_{\alpha }B_{\alpha }^{(n)}\otimes S_{\alpha }$, $%
h^{(n)}$ is closely related to a bound on $H_{e}^{(n)}$.

We consider the analog of case (i) in PDD, i.e., let $J<\beta $. Then, after
recursively applying Eq.~(\ref{eq:bx}) and the inequality $\left\Vert
\lbrack A,B]\right\Vert \leq 2\left\Vert A\right\Vert \left\Vert
B\right\Vert $ (valid for bounded operators $A$ and $B$ \cite{Bhatia:book}),
we have
\begin{eqnarray}
\left\Vert B_{X}^{(n)}\right\Vert &\leq & \prod_{i=0}^{n-1}(2\tau _{i}\beta
)\left\Vert B_{X}^{(0)}\right\Vert = J\prod_{i=0}^{n-1}(2\times 4^{i}\tau
_{0}\beta )  \notag \\
&=& 2^{n^{2}}(\beta \tau _{0})^{n}J.
\end{eqnarray}
Because of the factor of $1/2$ in $B_{Y}^{(n)}$ [compare Eqs.~(\ref{eq:bx}),(%
\ref{eq:by})] we also have:
\begin{equation}
\left\Vert B_{Y}^{(n)}\right\Vert \leq \left\Vert B_{X}^{(n)}\right\Vert
\hspace{1cm}\text{for sufficiently large }n\text{. }
\end{equation}%
Therefore, at the final concatenation level $n=n_{f}$
\begin{equation}
h^{{(n_{f})}}\leq 2^{n_{f}^{2}}(\beta \tau _{0})^{n_{f}}J.  \label{eq:hi}
\end{equation}%
The total duration $T$ is given in terms of $\tau _{0}$ and $n_{f}$ as:
\begin{equation}
T=\tau _{n_{f}}=4^{n_{f}}\tau _{0}=N\tau _{0}.  \label{eq:T}
\end{equation}

Let us now give the connection between the fidelity and the error phase. The
propagator corresponding to the whole sequence is given by $\exp\left[-i\tau
_{n_{f}}H_{e}^{(n_{f})}\right]$, and the ideal evolution of the system is
given by the identity operator. If fidelity is measured as the
\textquotedblleft state overlap between the ideal and the decoupled
evolution\textquotedblright , we can write \cite{Terhal:04}:
\begin{equation}
f\approx 1-\left\Vert T\underline{H}^{(n_{f})}\right\Vert ^{2}=1-\left( \tau
_{n_{f}}h^{(n_{f})}\right) ^{2}=1-\Phi _{\text{CDD}}^{2},  \label{eq:fid}
\end{equation}%
where $\underline{H}$ refers to the system-traceless part of $H$.

Generally, a different universal DD pulse sequence as the basic cycle of
concatenation will modify Eqs.~(\ref{eq:bx})-(\ref{eq:by}) but the
(asymptotic) form of Eq.~(\ref{eq:hi}) remains the same. The overall error
phase $\Phi _{\text{CDD}}=Th^{(n_{f})}$ can be bounded from above using
Eqs.~(\ref{eq:hi}),(\ref{eq:T}):
\begin{equation}
\Phi _{\text{CDD}}\leq (\beta T/N^{1/2})^{\log _{4}\!N}(JT).
\label{eq:PhiCDD}
\end{equation}%
We expect the above bound to be satisfied provided the convergence condition
$\beta T/N^{1/2}<1$, i.e., $\beta <(T\tau _{0})^{-1/2}$, is satisfied.
However, this condition is less strict than the condition appearing in
Appendix~\ref{appA}, for the truncation of the Magnus expansion:$\ \beta \ll
1/T$ [Eq.~(\ref{eq:taunb})].

For comparison, we also estimate $\Phi _{\text{PDD}}=Th^{(1)}$, i.e., simply
take $n_f=1$ in Eq.~(\ref{eq:hi}):
\begin{equation}
\Phi _{\text{PDD}}=2(\beta \tau _{0})(JT)=2(\beta T/N)(JT).
\label{eq:PhiPDD2}
\end{equation}%
Indeed, this agrees with Eq.~(\ref{eq:PhiPDD}). Note that for $N=4$, and
taking the equality signs in Eq.~(\ref{eq:PhiCDD}), we have as expected $%
\Phi _{\text{CDD}}=\Phi _{\mathrm{PDD}}$. Now, we may conclude, by comparing
Eqs.~(\ref{eq:PhiCDD}) and (\ref{eq:PhiPDD2}), that when $\beta <(T\tau
_{0})^{-1/2} $, $\Phi _{\text{CDD}}$ converges quickly to zero as the number
of pulses increases, while no such convergence is observed for PDD over the
same total sequence duration. In particular, while the fidelity gain in PDD
scales with the number of pulses $N$, it scales with $(N^{1/2}/c)^{\log
_{4}\!N}$($\gg N$ for $N\gg 1$) for CDD, where $c=\beta T$ is a small
constant regulated by the actual convergence domain. On the other hand, this
convergence domain puts a physical upper limit on the number of
concatenation levels, imposed by $\beta <(T\tau _{0})^{-1/2}$ or the
stricter $\beta \ll 1/T$ ($c\ll 1$).

Another way to compare CDD and PDD is as follows. By fixing the value of $c $
and $\beta $, we can back out an upper concatenation level: $n_{f}^{\max
}=-\log _{4}({\beta \tau _{0}}/{c})$. Inserting this into Eq.~(\ref%
{eq:PhiCDD}) we have:
\begin{equation}
\Phi _{\text{CDD}}\leq (c\beta \tau _{0})^{-\frac{1}{2}\log _{4}\frac{\beta
\tau _{0}}{c}}(JT).
\end{equation}%
We can now compare the CDD and PDD bounds:
\begin{equation}
\frac{\Phi _{\text{CDD}}}{\Phi _{\text{PDD}}}\leq \frac{(c\beta \tau _{0})^{-%
\frac{1}{2}\log _{4}\frac{\beta \tau _{0}}{c}}}{2(\beta \tau _{0})}\overset{%
\beta \tau _{0}\rightarrow 0}{\longrightarrow }0,  \label{eq:ratio}
\end{equation}%
\emph{which serves to show that CDD is indeed superior to PDD in the
(relevant) limit of small} $\beta \tau _{0}$. This key result was first
reported in our previous study \cite{KhodjastehLidar:04}, without a full
proof.

When the dynamics is dominated by direct system-environment coupling,
namely  $\beta \ll J$ [case (ii)], we find that the third order Magnus
term dominates the effective pure bath term. We find that the
effective coupling is then bounded by:
\begin{equation}
h^{{(n_{f})}}\leq c^{n_{f}^{2}}(\max(\beta^\prime,\beta) \tau _{0})^{n_{f}}J^4
\text{\hspace{.5cm} for $n_f>2$} \label{eq:hi-nob}
\end{equation}
where $c=O(1)$ and
\begin{equation}
  \beta^\prime=O(\tau_0^2 J^3)
  \label{beta'}
\end{equation}
is an effective pure-bath term that arises in the
3rd order Magnus expansion and kicks in at the second level of concatenation. The asymptotic
behavior in the effectiveness of dynamical decoupling is thus the same
as case (i) [compare to Eq.~(\ref{eq:hi})], however, due to the dependence on a higher power of
$J$, we see that this is a more favorable scenario for CDD.

Any universal decoupling sequence (e.g., higher order sequences) can be concatenated and our
analysis still applies. When this is done, Eq.~(\ref{eq:PhiCDD}) becomes
\begin{align}
\Phi_{\text{CDD}}\le \Phi_0 (\alpha N^{-a})^{\log N}N^b \label{eq:cdd-gen}
\end{align}
where $\Phi_0$ is the error phase for a free evolution of the system for time $T$. The parameter
$\alpha$ is generally bounded by a power of $\norm{H_e}T$ and is
required to be small [it is analogous to $\beta T$ in Eq.~(\ref{eq:PhiCDD})]. The
parameters $a$ and $b$ are $O(1)$. The parameters $\alpha,a,b$ all
depend on the basic decoupling cycle used and are all positive.

\subsection{Finite Width}

\label{finitewidth}

We can analyze the finite-width pulse CDD procedure by using the pulse error
unitary operators introduced in subsection~\ref{finite-width}. The
systematic error associated with these pulses at each level leads naturally
to a decoupling error, and is corrected at the next level of concatenation.
For convergence, we require the coupling strength to shrink as a function of
concatenation level. We obtain the following condition for convergence of
the concatenation procedure for rectangular pulses, derived in Appendix~\ref%
{appB}:

\begin{equation}
c^{\prime }\tau _{0}\beta +d^{\prime }\frac{\delta }{\tau _{0}}<1\text{.}
\label{eq:widthcond}
\end{equation}%
where $c^{\prime }$ and $d^{\prime }$ are numeric factors of $O(1)$. This
inequality is a special case of Ineq.~(\ref{eq:taunb'}), so the latter is
already sufficient to guarantee convergence. As mentioned there, the
finite-width convergence condition implies an optimal pulse interval $\tau
_{0}=\sqrt{d^{\prime }\delta /c^{\prime }\beta }$, for given $\beta $ and a
fixed minimal pulse width $\delta $. We reiterate that we have required $%
\delta \ll \tau _{0}$, but we expect this requirement to be inessential as
evident from the exact numerical simulations reported in \cite%
{KhodjastehLidar:04}. For example when a technological lower limit on the
smallest switching times was used, we obtained a \textquotedblleft
closed-pack sequence\textquotedblright\ with $\delta \approx \tau _{\text{min%
}}$ that still provides significant decoupling.

Generally, errors associated with non-zero pulse widths are a combination of
systematic and random pulse errors. By construction, the recursive nature of
CDD tolerates a significant level of systematic pulse errors, since the
decoupling error at each level is \textquotedblleft cleaned
up\textquotedblright\ at the next level of decoupling. Random pulse errors
are tolerated to some extent as well \cite{KhodjastehLidar:04}; this is
reminiscent of randomized decoupling techniques where pulses implementing
random unitary operators on the system are utilized for decoupling. In fact
randomized decoupling techniques have been shown to be efficient in the
limit of fast time-varying/fluctuating system-bath Hamiltonians for which
the deterministic methods (PDD/CDD) are relatively ineffective \cite%
{Santos:06}.

\subsection{Example: Decoupling in Spin Quantum Dots}

In this subsection we apply our analysis to the case of an electron spin in
a quantum dot coupled via the hyperfine interaction to a bath of nuclei. The
Hamiltonian for the interaction of an electron with spin $S$ (the qubit),
confined in a semiconductor quantum dot, with a collection of nuclear spins $%
I_{n}$, is given by:
\begin{eqnarray}
H &=&H_{S}+H_{SB}+H_{B}  \notag \\
&=&\Omega Z+\sum_{n}A_{n}I_{n}^{z}\otimes Z+\sum_{n<m}B_{nm}[I_{n}I_{m}]
\end{eqnarray}%
where $A_{n}$ and $B_{nm}$ are coupling constants and $B_{nm}[I_{n}I_{m}]$
is a shorthand for the diagonal and off-diagonal dipolar coupling terms
among the nuclei. The intra-nuclear dipolar coupling is short-ranged ($%
B_{nm}\propto r_{nm}^{-3}$). The hyperfine interaction between the electron
spin and the nuclei is a Fermi contact interaction, whose magnitude is given
by $A_{n}$, which is proportional to the electronic wave function magnitude
at the position of nucleus $n$. The parameters relevant to our bounds are
given by $J=O(I\sum_{n}A_{n})$ and $\beta =O(I^{2}\sum_{n<m}B_{nm})$, where $%
I(=O(1))$ denotes the nuclear spin. The number of nuclei within the radius of the electron
wavefunction is of the order of $10^{5}$ and for GaAs we have $\beta
\sim 10$kHz, and $J\sim 1$MHz \cite{merkulov:205309, Coish:06}. Let us
define an {\em error rate} as $e\equiv \Phi /T$ for an 
evolution of length $T$ with an error phase $\Phi $. The uncorrected (pulse-free) error rate
$e_{0}=\Phi /T$ is $O(J)\sim 1$MHz. Let us take our inter-pulse
interval $\tau _{0}=T/N$, to be 
smaller than $1\mu \text{s}$. At any pulse rate faster than this, PDD provides improvement:
$e_{\text{PDD}}=J^{2}\tau _{0}\lesssim 1\text{MHz}$
[Eq.~(\ref{eq:PhiPDD})]. Note that this error 
rate is fixed and applying more pulses at this rate only retains it,
resulting in an absolute error 
that grows linearly with time. For a detailed
example of this improvement see Ref. \cite{deSousa:06} where a
Carr-Purcell-Meiboom-Gill (CPMG) \cite{Carr:54,Meiboom:58} sequence is
used to decouple 
the electron spin from the nuclei (we give a concatenated version of the
CPMG sequence in Appendix~\ref{appC}).\footnote{The CPMG sequence is a simple dynamical decoupling
sequence consisting of fixed periodic spin flips corresponding to $X$
pulses in our notation. See Appendix~\ref{appC} for a detailed discussion.} In a system-specific
example such as that of Ref.~\cite{deSousa:06}, many of our conservative
bounds on the commutators appearing in the effective Hamiltonian can be
improved.

The error rate for CDD, $e_{\text{CDD}}$ is given by dividing
Eq.~(\ref{eq:PhiCDD}) by $T$ (or more generally by
Eq.~(\ref{eq:cdd-gen})), which decreases super-polynomially with the
number of pulses used. Note 
that, while initially we start in a regime where $\beta \ll J$ for low
levels of concatenation, the 
effective (renormalized) value of $J$ quickly decreases with
increasing concatenation level, and we 
are in a limit where $\beta$ (or the effective $\beta^{\prime}$
[Eq.~(\ref{beta'})]) dominates through the undecoupled
error term. Finally, we note that as long as pulse widths that are
negligible with respect to our 
pulse intervals are used, the condition for convergence of
concatenation [Eq. (\ref{eq:widthcond})] is satisfied in our example.
\begin{figure}
\includegraphics[width=\columnwidth]{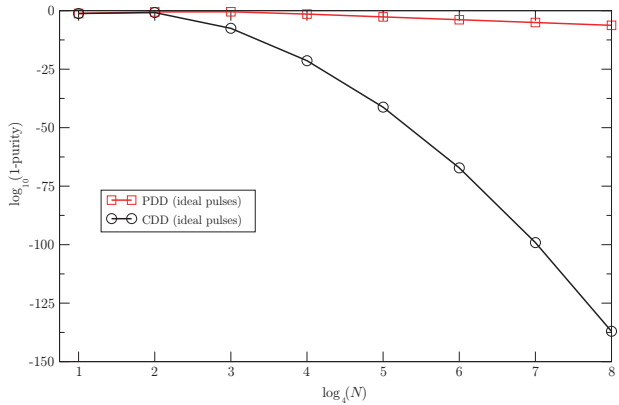}
\caption{Simulation of PDD (square) and CDD (circles) for
  concatenation levels 1 to 8. Results shown are for one minus
  purity of the system qubit as a function of the number of pulses.
  The pulses used are ideal. Note the fantastically high purity
  achieved with CDD. In the case of PDD the horizontal axis gives
  $\log_4$ of the
  total number of pulses, while for CDD the numbers on the horizontal
  axis denote the concatenation level.}
\label{fig:1}
\end{figure}%

We performed a numerically exact simulation for comparing different
pulse sequences for a qubit coupled to a small environment. We
consider a spin chain of length $N$, where the central spin is from a
different species (the electron in the dot versus the surrounding
nuclei) and acts as the system qubit. The (undesired) couplings
between the spins is given by Heisenberg interaction terms, so that
the coupling Hamiltonians between two spins $i$ and $j$ at a distance
$d$ (measured in units of the lattice constant) is given by
$\frac{c}{2^d}(X_iX_j+Y_iY_j+Z_iZ_j)$, where $c$ depends on the
species of the spins $i$ and $j$. We fix the coupling strengths and
the number of spins so that the coupling strengths roughly correspond
to GaAs our example: $J=1\text{MHz}$ and $\beta=10\text{KHz}$. The
effective errors at the end of the concatenated cycles are so low
(purity loss of around $10^{-100}$ -- see Fig.~\ref{fig:1}) that we are forced to use extremely
high precision linear algebra that significantly burdens simulation
performance. To get practical results, we argue based on Lieb-Robinson
bounds (see \cite{osborne:157202}) that within the duration of the
simulation, enlarging the spin chain will not modify the results. This
was verified independently for a test case where there was no
significant qualitative difference between $N=3,5,7$. Therefore the
results presented here are for $N=3$ and they capture the essence of
our estimates.
\begin{figure}
\includegraphics[width=\columnwidth]{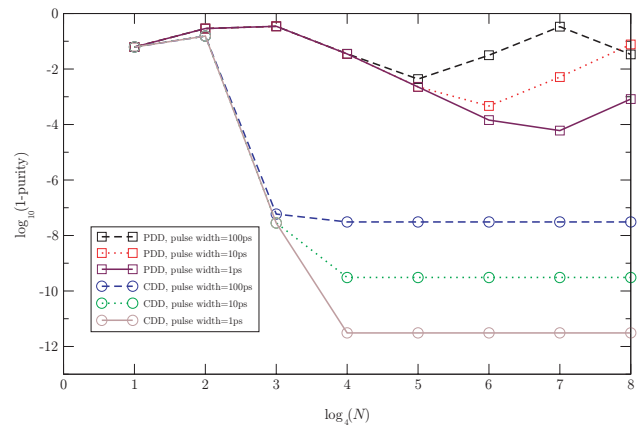}
\caption{Simulation of PDD (squares) and CDD (circles) for
concatenation levels 1 to 8 with non-ideal pulses. The pulse widths
are set at 1ps, 10ps, and 100ps. Axes as in Fig.~\protect\ref{fig:1}.}
\label{fig:2}
\end{figure}
We fix the overall duration of the pulse sequence at $10^{-5}\text{s}$
so that $\beta T<1$ [see Eq.  (\ref{eq:taunb'})]. This allows us to
concentrate on the inter-pulse period and pulse-widths that are the
main technological challenge for decoupling. Higher concatenation
levels thus correspond to $\log(\text{number of pulses})$ [or to $\log(1/\text{inter-pulse period})$]. In Fig.~\ref{fig:1}, we compare CDD and PDD purities for ideal pulses at
various levels of concatenation. The vertical axis displays the loss
of purity of the system qubit, i.e., $1-\textrm{Tr}[(\rho_S)^2]$,
where $\rho_S = \textrm{Tr}_B (\rho)$, and $\rho$ is the joint
system-bath density matrix at the final time $T$. The higher order data for PDD are
obtained by simply repeating the basic sequence and shrinking the
pulse-interval. This is necessary to get the long-term behavior of
decoupling. The graphs show a progressive improvement in purity of the
initial qubit state, $\frac{\ket{0}+\ket{1}}{\sqrt{2}}$, as a function
of concatenation level. The environment part of the spin chain is
initialized in a thermal state at a temperature of $1$K.
In Fig.~\ref{fig:2} we have depicted the effect of realistic pulse
widths on decoupling. We note that in the case of PDD the performance
of the pulse sequence deteriorates with increasing pulse width (as
expected), but also (after an initial improvement, as in the ideal
pulse case) as a function of
the number of pulses used. This latter deterioration is somewhat
surprising, and is due to
the pulse width errors that simply accumulate over time. The point of
deterioration shifts to the right as the pulse width is made smaller,
as expected. The improvement seen for the $100$ps case at
$\log_4(N)=7$ can be understood as being due to the essentially
random-walk-on-a-circle-like nature of the accumulated errors, which
will occassionally result in a recurrence. Concatenated sequences
are naturally robust against these errors, as can be seen clearly in
the figure, where CDD results in a saturation of the purity. The
asymptotic purity level is roughly equal to the square of the per-pulse-error $\delta
J$ (recall subsection~\ref{finite-width}). For example, with
$\delta=1$ps and $J=1$MHz we find $(\delta J)^2=10^{-12}$, in
agreement with Fig.~\ref{fig:2}.
While it is impossible to
go to error rates below per-pulse-error, with concatenation we are
able to maintain the error rate at this minimum.

\section{Higher Order (Trotter-Suzuki) Universal Decoupling}

\label{TSD}

Suppose a universal dynamical decoupling sequence is known, i.e.,
essentially a series of unitarily transformed Hamiltonians $%
H_{j}=P_{j}H_{e}P_{j}^{\dag }$ such that $\sum_{j}P_{j}H_{e}P_{j}^{\dag
}=B\otimes I_{S}$ for some environment operator $B$. As we saw previously,
the sequential application of propagators generated by these Hamiltonians
acts trivially on the system only up to the first order in $\Vert B_{\alpha
}\Vert \tau _{0}$ where $\tau _{0}$ is a typical free-evolution period
duration. Trotter-Suzuki expansion allows us to construct a sequence of
these propagators that act trivially on the system up to any order $n$.
Suppose $\{\epsilon A_{j}\}_{j=1}^{k}$ are dimensionless Hermitian operators
such that $\sum A_{j}=A$ and $\epsilon $ is a small parameter.
Trotter-Suzuki expansion allows us to find a sequence of indices $%
\{n_{i}\}_{i=1}^{N}$ ($n_{i}\in \{1\ldots k\}$) and real numbers $%
\{c_{i}\}_{i=1}^{N}$ such that
\begin{equation}
e^{\epsilon c_{1}A_{n_{1}}}e^{{\epsilon c_{2}A_{n_{2}}}}\ldots e^{{\epsilon
c_{N}A_{n_{N}}}}=e^{\epsilon A+O(\epsilon ^{n})}  \label{eq:suz}
\end{equation}%
The number $N$ of propagators required, scales with $e^{n^{2}}$. It is worth
emphasizing that while numerous, the sequence of coefficients $c_{i}$ and
indices $n_{i}$ can be constructed recursively \cite{Suzuki:76, Hatano:05}.

For practical purposes it is advantageous to use a variation on this
expansion in which the coefficients $c_{i}$ are rational numbers. Once the
sequence of propagators is known we can construct the Trotter-Suzuki
decoupling sequence (TSDS) accordingly: Set up $A_{j}=iH_{j}$ and $%
A=iB\otimes I$ based on the universal DD cycle and employ the following DD
sequence using the smallest pulse switching time $\tau _{\text{min}}$
available:
\begin{equation}
P_{n_{1}}\mathtt{f}[c_{1}\tau _{\text{min}}]P_{n_{1}}^{\dag }P_{n_{2}}%
\mathtt{f}[c_{2}\tau _{\text{min}}]P_{n_{2}}^{\dag }\ldots P_{n_{N}}\mathtt{f%
}[c_{N}\tau _{\text{min}}]P_{n_{N}}^{\dag },  \label{eq:TSD}
\end{equation}%
where $\mathtt{f}[\tau ]$ denotes a free evolution period of duration $\tau $%
. By dimensional analysis, the error in Eq.~(\ref{eq:suz}), $\epsilon ^{n}$,
translates into left-over terms asymptotically bounded by products of
operators $B_{\alpha }$ and is thus bounded by $O[(\tau _{\mathrm{min}}\max {%
\Vert B_{\alpha }\Vert })^{n}]$. We can thus asymptotically estimate the
error phase as:
\begin{eqnarray}
\Phi _{\text{TSDS}} &=&O\left[ \left( \tau _{\text{min}}\max \{\Vert
B_{\alpha }\Vert \}\right) ^{n}\right]  \notag \\
&=&O\left[ \left( T\max \{\Vert B_{\alpha }\Vert \}/N\right) ^{\sqrt{\log N}}%
\right] ,
\end{eqnarray}

A serious problem is that negative $c_{i}$'s routinely appear in the TSDS
(except for $n=3$). Clearly, this presents a major problem since one cannot
have negative times in the free evolution segments in Eq.~(\ref{eq:TSD}).
For this reason, as presented the TSDS is not physically implementable. A
solution for this problem is not yet available to us, but does not seem
impossible.

Table (\ref{tab1}) summarizes the asymptotic performance of PDD, CDD and
TSDS in the limit of large number of pulses and the two regimes of $J<\beta $
and $\beta \ll J$.
\begin{table}[tbp]
\begin{tabular}{|c|c|c|}
\hline
DD Scheme & $\Phi$ for $J<\beta$ & $\Phi$ for $\beta \ll J $\\
&  & \\ \hline Periodic DD & $T^2\beta J/N$ & $T^2 J^2/N$\\ \hline
Concatenated DD & $(\beta T/N^{1/2})^{\log_4\!N}(JT)$ & $N{({JT}/{N})}^{N^{\log_4 (5/2)}}$\\ \hline
Trotter-Suzuki & $(\beta T/N)^{\sqrt{\log_4\!N}}$ & $(J T/N)^{\sqrt{\log_4\!N%
}}$\\ \hline
\end{tabular}%
\caption{Asymptotic comparison of the error phase $\Phi $ for deterministic
decoupling schemes. $T$ is the total experiment duration, $N$ the total
number of pulses, $J$ and $\protect\beta $ are defined in Eq.~(\protect\ref%
{eq:Jb}).}
\label{tab1}
\end{table}

The TSDS performs remarkably well and is oblivious to the distinction
between pure-bath and system-bath dynamics. Nonetheless, its performance is
sensitive to small errors in pulse operation and the switching times that
need to be precisely set (not required in PDD and CDD and randomized
decoupling schemes). Thus it may not be a robust alternative to CDD, in
spite of its superior convergence properties, but it may be used as a basic
dynamical decoupling sequence at the base of a new concatenated pulse
sequence. This is particularly useful with the TSDS at $n=3$.

Finally, we note that the Trotter-Suzuki expansion was also recently used by
Brown \textit{et al.} in a study of arbitrarily accurate composite pulse
sequences \cite{Brown:04}. There the goal was to overcome systematic errors
in the system control Hamiltonian, without considering decoherence.

\section{Decoupling with Very Narrow Pulses Cannot Increase Error Norms}

\label{Thompson}

In this section we return to DD\ with ideal, zero width pulses, and argue
that such DD sequences can never effectively strengthen the undesired terms
and cannot cause extra errors. We then argue that even with finite-width,
but sufficiently narrow pulses, error norms cannot increase under DD. These
results are of independent interest and apply to any pulse-based error
suppression strategy, including closed-loop quantum error correction.

Consider a sequence of ideal unitary operations $P_{i}$ applied to a quantum
system (measurements can also be included by enlarging the Hilbert space),
and suppose a sequence of intervals $\tau _{i}$ separates these operations,
such that the pulse sequence is given by: $P_{1}{\small \mathtt{f}_{\tau
_{1}}}P_{2}{\small \ldots \mathtt{f}_{\tau _{n-1}}}P_{n}$. In the absence of
environmental couplings, the overall effect of the pulse sequence would be
given by $Q_{n}=P_{1}\ldots P_{n}$ -- the \textquotedblleft ideal
operation\textquotedblright . But in the presence of the environmental
couplings in $H_{e}$, the overall propagator $U$ is modified by error terms.
Letting $Q_{i}=P_{1}\ldots P_{i}$, we can define an effective unitary error
operator corresponding to the whole sequence in the following manner:
\begin{align}
U& =P_{1}e^{-i\tau _{1}H_{e}}P_{2}e^{-i\tau _{2}H_{e}}P_{3}\ldots
P_{n-1}e^{-i\tau _{n}H_{e}}P_{n}  \notag \\
& =Q_{1}e^{-i\tau _{1}H_{e}}Q_{1}^{\dag }Q_{2}e^{-i\tau
_{2}H_{e}}Q_{2}^{\dag }\ldots Q_{n-1}e^{-i\tau _{n-1}}Q_{n-1}^{\dag }Q_{n}
\notag \\
& =e^{-i\tau _{1}Q_{1}H_{e}Q_{1}^{\dag }}e^{-i\tau _{2}Q_{2}H_{e}Q_{2}^{\dag
}}\ldots e^{-i\tau _{1}Q_{n-1}H_{e}Q_{n-1}^{\dag }}Q_{n}  \notag \\
& =:\exp (-iTH_{e}^{\prime })P_{1}\ldots P_{n}  \label{eq:efer}
\end{align}%
Note that in the last line we have defined the overall duration $T=\sum \tau
_{i}$, and isolated the effective error Hamiltonian $H_{e}^{\prime }$ from
the desired, ideal unitary operation $Q_{n}=P_{1}\ldots P_{n}$.

As before, we use a unitary operator norm $\Vert .\Vert $ to compare the
strength of the error Hamiltonians $H_{e}$ and $H_{e}^{\prime }$. Unitary
operator norms \cite{Bhatia:book}, such as the absolute difference between
the largest and smallest eigenvalues, are invariant under unitary
transformations (for unitary $U$, $\Vert UAU^{\dag }\Vert =\Vert A\Vert $)
and can be used as measures of fidelity errors in Hamiltonian error
correction theory \cite{Terhal:04}. We now use the following existential
theorem due to Thompson \cite{Thompson:86,Childs:03}: Let $A_{1}$, $A_{2}$
be Hermitian matrices. Then there exist unitaries $U_{1}$ and $U_{2}$ and a
Hermitian matrix $A$ such that
\begin{equation*}
e^{iA_{1}}e^{iA_{2}}=e^{iA};\hspace{5mm}A=U_{1}A_{1}U_{1}^{\dag
}+U_{2}A_{2}U_{2}^{\dag }.
\end{equation*}%
This theorem can be extended (induction on the number of exponentials) to
products involving more than two matrix exponentials: $e^{iA_{1}}\cdots
e^{iA_{n}}=e^{iA}$. Using Eq.~(\ref{eq:efer}) and Thompson's theorem we
have: $TH_{e}^{\prime }=\sum_{i}\tau _{i}U_{i}Q_{i}H_{e}Q_{i}^{\dag
}U_{i}^{\dag }$, and we have the following inequality for the norm of $%
H_{e}^{\prime }$:
\begin{align}
\Vert H_{e}^{\prime }\Vert &=\frac{1}{T}\Vert \sum_{i}\tau
_{i}U_{i}Q_{i}H_{e}Q_{i}^{\dag }U_{i}^{\dag }\Vert  \notag \\
&\leq \frac{1}{T}\sum_{i}\tau _{i}\Vert U_{i}Q_{i}H_{e}Q_{i}^{\dag
}U_{i}^{\dag }\Vert =\Vert H_{e}\Vert  \label{eq:He'}
\end{align}%
Thus the norm of the effective error Hamiltonian does not increase under the
action of ideal unitary operators.

In the case of non-ideal pulses carrying systematic errors per pulse (e.g.,
due to finite pulse-width), we can model the pulse error as a unitary error
operator immediately preceding the ideal pulse. For single-qubit pulse
errors due to finite pulse widths, one can again show that $\Vert
H_{e}^{\prime }\Vert \leq \Vert H_{e}\Vert $ provided the pulse widths are
small enough \cite{KhodjastehLidar:inprep}. This allows us to use Thompson's
theorem to show that our argument in this section also applies to pulses of
sufficiently narrow width on a single qubit. We expect this argument to
apply to multi-qubit near-perfect operators in the presence of a bounded
bath.

As a special case, this argument applies to dynamical decoupling. While
positively reassuring that with ideal pulses the undesired couplings do not
increase in strength, the present argument does not quantify the efficiency
of dynamical decoupling. For this purpose we employed, above, approximations
based on the Magnus expansion.

The same argument applies in a quantum error correction codes setting \cite{Knill:97a}, in particular in non-Markovian fault-tolerance theory \cite{Aharonov:05,Aliferis:05,Terhal:04}, where a \textquotedblleft time-resolved fault
path\textquotedblright\ expansion has recently been used to decompose the action of errors in the
course of a general evolution. Our argument can be used to further rationalize such expansions
based on the fact that errors are well-behaved (in the sense of $\Vert H_{e}^{\prime }\Vert \leq
\Vert H_{e}\Vert $) in the non-Markovian regime.

\section{Summary and Discussion}

\label{summary}

Dynamical decoupling (DD) cannot be exactly analyzed without concrete
reference to the details of the system-environment coupling, but an abstract
picture of the interaction in terms of bounded environment operators -- as
pursued here -- can yield useful performance estimates. Within this
framework, we have provided an analytic estimate of the leading order
decoupling error associated with the basic universal decoupling cycle for a
qubit. We have analyzed and compared the performance of periodic DD (PDD)
and concatenated DD (CDD) schemes. We have provided detailed calculations
supporting the conclusion reported in \cite{KhodjastehLidar:04}, that CDD
significantly outperforms PDD within practical boundaries of pulse parameter
space. We have distinguished between two different limiting cases of fast
versus slow environment dynamics. Fast dynamics of the environment limits
the performance of higher order deterministic dynamical decoupling. This can
be understood from the interaction picture, where the system-bath
interaction Hamiltonian is fast fluctuating. Slow bath dynamics, on the
other hand, can be exploited by CDD to result in super-exponential
decoupling using an exponential number of pulses. Table~\ref{tab1} provides
a convenient summary of the relative performance of PDD and CDD, as well as
the new Trotter-Suzuki based pulse sequence we have introduced.

Our discussion was based on a pulsed control mode, but it is known that
higher fidelities are possible via fine-tuned navigation of the control
Hamiltonian \cite{Viola:02}. This direction can be especially useful when
decoupling methods are to be used not for quantum state preservation but for
performing an error-corrected quantum evolution. Dynamically error corrected
evolution can also be achieved via a hybrid decoupling error correction
method \cite{ByrdLidar:01a,ByrdLidar:02a,KhodjastehLidar:03}: Consider a
stabilizer quantum error correcting code characterized by a stabilizer group
of Pauli operators \cite{Calderbank:97,Gottesman:96}. Assume that this code
corrects all single qubit errors, which means that each single qubit error
anticommutes with at least one stabilizer generator. This anti-commutation
condition translates into time reversal of the error, when exponentiated.
Thus it is possible to use the generator set of the stabilizer to form a
universal decoupling sequence for decoupling single qubit error terms on the
code space. In this way decoupling integrates seamlessly with the encoded
quantum operations on the code space generated by Hamiltonians written as
the sum of normalizer elements of the code. This also allows us to perform
quantum error detection and recovery within a hybrid decoupling-error
correction setting, thus allowing the correction of errors in both the
Markovian and the non-Markovian regimes \cite{KhodjastehLidar:inprep}.

The open-loop approach of DD appears at first sight to be conceptually and
practically very different from the method of closed-loop quantum error
correcting codes \cite{Steane:99}. However, recent results on the theory of
non-Markovian fault tolerant quantum error correction (FTQEC) \cite%
{Terhal:04,Aliferis:05,Aharonov:05} suggest that the error bounds applicable
in the fault tolerant, concatenated version of DD are in fact very similar
to the bounds relevant to FTQEC. Specifically, in both CDD and FTQEC it is
essential that the system-bath interaction Hamiltonian is norm-bounded (see
Refs.~\cite{AlickiLidarZanardi:05,Alicki:07} for a critique of this assumption). In
light of the much smaller degree of overhead involved in CDD, this suggests
that in the non-Markovian regime one can profit significantly by
incorporating CDD into a closed-loop QECC procedure. CDD can then remove the
leading order bath-induced errors, while the QECC procedure can target
primarily the random control errors for which CDD offers only limited
protection.

We note that it is possible to view the CDD procedure as a discrete time
\emph{dynamical system}, whose ideal fixed point is a vanishing system-bath
interaction. In this manner it should be possible to characterize the region
of correctable errors using tools from the analysis of fixed points, and to
incorporate perturbations of the pulse sequence. An analysis of CDD from
this perspective may well be a fruitful endeavor. Indeed, there exists a
dynamical maps approach to concatenated quantum error correction, which has
proven to be very convenient in the analysis of the fault tolerance
threshold \cite{Rahn:02,Fern:06}.

As a final comment we should emphasize that the ability to perform
arbitrarily precise Hamiltonian control on the system Hamiltonian assumes a
classical control mechanism, while every control system is really quantum in
nature. Thus besides technological constraints, fundamental quantum
fluctuations may limit the performance of DD as well, since perfect
classical control simply does not exist. A systematic characterization of
bounds on the fidelity of feedback-free error correction schemes such as DD,
imposed by fundamental quantum fluctuations, is still an important open
question.

\section*{Acknowledgements}

We would like to acknowledge helpful discussions with W. Yao, M. A. Nielsen,
L. Viola, and P. Zanardi. This work was supported under grants NSF
CCF-0523675 and ARO W911NF-05-1-0440.

\appendix

\section{Validity of the Magnus Expansion}

\label{appA}

We analyze the convergence domain of approximations and the dynamical
renormalization process, specifically as it applies to CDD. This recursive
renormalization can be written as:
\begin{equation}
U^{(n+1)}:=\exp (-i\tau _{n+1}H_{e}^{(n+1)})=\prod\limits_{j=1}^{4}\exp
(-i\tau _{n}H_{j}^{(n)})
\end{equation}%
where $\tau _{n+1}=4\tau _{n}$, i.e.,
\begin{equation}
\tau _{n}=4^{n}\tau _{0},
\end{equation}%
and where
\begin{widetext}
\begin{align}
H_{1}^{(n)} &=B_{0}^{(n)}\otimes I+B_{X}^{(n)}\otimes X+B_{Y}^{(n)}\otimes
Y+B_{Z}^{(n)}\otimes Z=IH_{e}^{(n)}I,  \notag \\
H_{2}^{(n)} &=B_{0}^{(n)}\otimes I+B_{X}^{(n)}\otimes X-B_{Y}^{(n)}\otimes
Y-B_{Z}^{(n)}\otimes Z=XH_{e}^{(n)}X,  \notag \\
H_{3}^{(n)} &=B_{0}^{(n)}\otimes I-B_{X}^{(n)}\otimes X+B_{Y}^{(n)}\otimes
Y-B_{Z}^{(n)}\otimes Z=YH_{e}^{(n)}Y,  \notag \\
H_{4}^{(n)} &=B_{0}^{(n)}\otimes I-B_{X}^{(n)}\otimes X-B_{Y}^{(n)}\otimes
Y+B_{Z}^{(n)}\otimes Z=ZH_{e}^{(n)}Z,
\end{align}
\end{widetext}
are the recursive generalization of Eqs.~(\ref{eq:Hi}). The Magnus expansion
of $U^{(n)}$ yields:
\begin{equation}
U^{(n)}=\exp (-i\tau _{n}H^{(n)})=\exp (\sum_{i=1}^{\infty }A_{i}^{(n)}),
\end{equation}
whence
\begin{equation}
\tau _{n}H^{(n)}=i[A_{1}^{(n)}+A_{2}^{(n)}]+\tau _{n}C^{(n)},
\end{equation}
where
\begin{align}
A_{1}^{(n)} &=-i\tau _{n-1}\sum_{i=1}^{4}H_{i}^{(n-1)},  \label{eq:A1n} \\
A_{2}^{(n)} &=-\frac{1}{2}\tau
_{n-1}^{2}\sum_{1=i<j=4}[H_{j}^{(n-1)},H_{i}^{(n-1)}],  \label{eq:A2n} \\
-i\tau _{n}C^{(n)} &=\sum_{i=3}^{\infty }A_{i}^{(n)}=\sum_{i=3}^{\infty
}m_{i}\tau _{n-1}^{i}\underset{i\text{ commutators}}{\underbrace{%
[H_{j}^{(n-1)},[H_{k}^{(n-1)},...]]}},  \label{eq:Cn}
\end{align}
where $m_{i}$ is a numerical factor determined by explicit computation of
the $i$th order Magnus expansion, and $C^{(n)}$ is an operator-valued
correction to the second order Magnus expansion.

We would like to find an approximation for the $B_{\alpha }^{(n)}$. To do so
we will first show that it is consistent to use the second order Magnus
expansion for $H_{e}^{(n)}$, in the sense that
\begin{equation}
H_{1}^{(n)}=H_{e}^{(n)}\approx \frac{i}{\tau _{n}}%
(A_{1}^{(n)}+A_{2}^{(n)})=:\sum_{\alpha =0,X,Y,Z}\tilde{B}_{\alpha
}^{(n)}\otimes S_{\alpha }\text{,}  \label{eq:He(n)}
\end{equation}%
where we can safely neglect $C^{(n)}$ [i.e., all $A_{i>2}^{(n)}$] \emph{as
long as}
\begin{equation}
\tau _{n}\beta \ll 1.  \label{eq:taunb}
\end{equation}
This is the recursive generalization of the result obtained above for $n=1$.
Then the $\tilde{B}_{\alpha }^{(n)}$ will be the desired approximation to $%
B_{\alpha }^{(n)}$. The proof is by induction. We will require the following
inequalities, satisfied for bounded operators $A$ and $B$ \cite{Bhatia:book}
:
\begin{eqnarray}
\left\Vert \lbrack A,B]\right\Vert &\leq &2\left\Vert A\right\Vert
\left\Vert B\right\Vert ,  \label{ineq:com} \\
\left\Vert AB\right\Vert &\leq &\left\Vert A\right\Vert \left\Vert
B\right\Vert ,  \label{ineq:mul} \\
\left\Vert A+B\right\Vert &\leq &\left\Vert A\right\Vert +\left\Vert
B\right\Vert .  \label{ineq:plu}
\end{eqnarray}

\begin{lemma}
The following relations hold:%
\begin{equation}
||C^{(n)}||\ll \frac{1}{\tau _{n}}||A_{2}^{(n)}||  \label{eq:C(n)}
\end{equation}%
and%
\begin{eqnarray}
||\tilde{B}_{\alpha }^{(n)}|| &=&O(\beta ),\quad \alpha =X,Y,Z
\label{eq:Ba(n)} \\
B_{0}^{(n)} &=&B_{0}.  \label{eq:B0(n)}
\end{eqnarray}
\end{lemma}

\begin{proof}
We prove the lemma by induction. Let us call Eq.~(\ref{eq:C(n)})
\textquotedblleft a($n$)\textquotedblright , Eq.~(\ref{eq:Ba(n)})
\textquotedblleft b$_{\alpha }$($n$)\textquotedblright , and Eq.~(\ref%
{eq:B0(n)}) \textquotedblleft b$_{0}$($n$)\textquotedblright . We have
already established the case a($1$) [Eq.~(\ref{eq:A2i})], and b$_{\alpha }$($%
1$) and b$_{0}$($1$) are based on our definitions and assumptions. We will
show that (1) b$_{\alpha }$($n-1$) \& b$_{0}$($n-1$) $\Rightarrow $ a($n$),
(2) b$_{0}$($n-1$) $\Rightarrow $ b$_{0}$($n$), and then (3) a($n$) \& b$%
_{\alpha } $($n-1$) $\Rightarrow $ b$_{\alpha}$($n$). Recall all along that
we have assumed $J<\beta $.
\end{proof}

(1) Proof of a($n$): We have, using Eqs.~(\ref{eq:A2n}),(\ref{eq:Cn}) and
inequality (\ref{ineq:com})
\begin{widetext}
\begin{align*}
\tau _{n}||C^{(n)}|| &\leq \sum_{i=3}^{\infty }m_{i}\tau _{n-1}^{i}||
\underset{i\text{ commutators}}{\underbrace{%
[H_{j}^{(n-1)},[H_{k}^{(n-1)},...]]}}|| =O\left(\sum_{i=3}^{\infty }\tau _{n-1}^{i-2}\underset{i-2\text{ terms}}{
\underbrace{(||H_{j}^{(n-1)}||\,||H_{k}^{(n-1)}||\cdots
)}}||A_{2}^{(n)}||\right)
\\
&\overset{\text{b}_{\alpha }\text{(}n-1\text{) }\&\text{ b}_{0}\text{(}n-1
\text{)}}{=}O\left(\sum_{i=3}^{\infty }\tau _{n-1}^{i-2}\beta
^{i-2}||A_{2}^{(n)}||\right) \overset{\tau _{n-1}\beta \ll 1}{=}O(\tau _{n-1}\beta ||A_{2}^{(n)}||) \ll||A_{2}^{(n)}||
\end{align*}

(2) Proof of b$_{0}$($n$):
\begin{align*}
  A_{1}^{(n)} &=-i\tau _{n-1}\sum_{i=1}^{4}H_{i}^{(n-1)} =-i4\tau
_{n-1}B_{0}^{(n-1)}\otimes I\overset{\text{b}_{0}\text{(}n-1\text{)}}{=}
-i\tau _{n}B_{0}\otimes I .
\end{align*}

(3) Proof of b$_{\alpha }$($n$): From Eq.~(\ref{eq:He(n)}) we have
\begin{align*}
H_{e}^{(n)} &=\sum_{\alpha =0,X,Y,Z}B_{\alpha }^{(n)}\otimes S_{\alpha }
\overset{\text{a(}n\text{)}}{\approx }\frac{i}{\tau _{n}}%
(A_{1}^{(n)}+A_{2}^{(n)}) =\sum_{\alpha =0,X,Y,Z}\tilde{B}_{\alpha
}^{(n)}\otimes S_{\alpha }.
\end{align*}
Since $A_{2}^{(n)}$ contains no pure-environment terms it determines the
part contributing to the sum over $\alpha =X,Y,Z$:
\begin{eqnarray*}
\norm{\hspace{-.5cm} &\sum_{\alpha =X,Y,Z}& \hspace{-.5cm} \tilde{B}_{\alpha }^{(n)}\otimes S_{\alpha }} =
\frac{1}{\tau _{n}}||A_{2}^{(n)}|| \overset{\text{Eq. (\ref{eq:A2n})}}{=}
\frac{1}{\tau _{n}}||\frac{1}{2}\tau
_{n-1}^{2}\sum_{1=i<j=4}[H_{j}^{(n-1)},H_{i}^{(n-1)}]||  \\
&\overset{\text{Ineq. (\ref{ineq:com})-(\ref{ineq:plu})}}{\leq}&\frac{\tau
_{n-1}^{2}}{4\tau
  _{n-1}}\sum_{1=i<j=4}||H_{j}^{(n-1)}||\,||H_{i}^{(n-1)}||\overset{\text{b}_{\alpha }\text{(}n-1\text{)}}{=} O(\tau _{n-1}J^{2})
<O(\tau _{n-1}\beta ^{2})<O(\beta ).
\end{eqnarray*}
Since the system operators all have $||S_{\alpha }||=O(1)$ and the
environment operators $\tilde{B}_{\alpha }^{(n)}$ all have similar norm, we
can conclude that, as required, $||\tilde{B}_{\alpha }^{(n)}||=O(\beta )$.

The upshot of this proof is the following:

\begin{corollary}
The recursive second-order Magnus expansion, Eq.~(\ref{eq:He(n)}), is a
valid approximation provided we assume $\tau _{n}\beta \ll 1$.
\end{corollary}

The condition $\tau _{n}\beta \ll 1$ of course puts a physical upper limit
on the number of levels of concatenation. Provided this condition is
satisfied, it follows that, schematically, we have
\begin{align}
H_{e},\tau _{0}& \mapsto H_{e}^{(1)}\approx \frac{i}{\tau _{1}}
(A_{1}^{(1)}+A_{2}^{(1)})=:\sum_{\alpha }\tilde{B}_{\alpha }^{(1)}\otimes
S_{\alpha },  \notag \\
H_{e}^{(1)},4\tau _{0}& \mapsto H_{e}^{(2)}\approx \frac{i}{\tau _{2}}
(A_{1}^{(2)}+A_{2}^{(2)})=:\sum_{\alpha }\tilde{B}_{\alpha }^{(2)}\otimes
S_{\alpha },  \notag \\
& \vdots  \notag \\
H_{e}^{(n-1)},4^{n-1}\tau _{0}& \mapsto H_{e}^{(n)}\approx \frac{i}{\tau
_{n} }(A_{1}^{(n)}+A_{2}^{(n)})=:\sum_{\alpha }\tilde{B}_{\alpha
}^{(n)}\otimes S_{\alpha }.  \label{eq:Hn}
\end{align}
\end{widetext}
Note that in the body of the paper, for notational simplicity we dropped the
tilde, with the convention being that we are only considering the
environment operators defined by the second-order Magnus expansion.

\section{Finite Pulse Width Analysis for CDD}

\label{appB}

For brevity define
\begin{equation}
\beta _{\alpha }^{(n)}:=\Vert B_{\alpha }^{(n)}\Vert ,\quad \beta _{\alpha
}:=\beta _{\alpha }^{(0)},\quad \beta :=\beta _{0},
\end{equation}%
and use Ineqs. (\ref{ineq:com})-(\ref{ineq:plu}) to reproduce the recursive
inequalities corresponding to $\beta _{\alpha }^{(n)}$ from Eqs.~(\ref%
{eq:B0(n)-fw})-(\ref{eq:BZ(n)-fw}):
\begin{align}
\beta _{0}^{(n)}& =\beta , \\
\beta _{X}^{(n)}& \leq 2(\tau _{n-1}-\delta )\beta \beta _{X}^{(n-1)}+\frac{%
\delta }{\tau _{n-1}}(\frac{1}{2}\beta _{X}+\frac{1}{\pi }\beta _{Y})
\label{eq:betaX(n)} \\
\beta _{Y}^{(n)}& \leq (\tau _{n-1}-\delta )\beta \beta _{Y}^{(n-1)}+(\tau
_{n-1}-2\delta )\beta _{X}^{(n-1)}\beta _{Z}^{(n-1)}  \notag \\
& \hspace{.5cm} +\frac{1}{\pi }\frac{\delta }{\tau _{n-1}}\beta _{Z}
\label{eq:betaY(n)} \\
\beta _{Z}^{(n)}& \leq \delta \left[ \frac{2}{\pi }\beta _{X}^{(n-1)}(\beta
_{Z}+\beta _{X})+\beta _{Y}^{(n-1)}\beta _{X}\right] +\frac{\delta }{\tau
_{n-1}}\beta _{Z}.  \label{eq:betaZ(n)}
\end{align}%
A necessary condition for convergence is $\beta _{\alpha }^{(1)}<\beta
_{\alpha }^{(0)}\equiv \beta _{\alpha }$ for $\alpha =X,Y,Z$. Let us define
constants $a,b$ such that%
\begin{equation}
\beta _{Y}=a\beta _{X},\quad \beta _{Z}=b\beta _{X}.
\end{equation}%
Then we have for the $n=1$ case of Eq.~(\ref{eq:betaX(n)})
\begin{equation}
\beta _{X}^{(1)}\leq 2(\tau _{0}-\delta )\beta \beta _{X}+\frac{\delta }{%
\tau _{0}}(\frac{1}{2}+\frac{a}{\pi })\beta _{X},
\end{equation}%
which must be smaller than $\beta _{X}$. We thus find the necessary
condition
\begin{equation}
2(\tau _{0}-\delta )\beta +\frac{\delta }{\tau _{0}}(\frac{1}{2}+\frac{a}{%
\pi })<1.  \label{eq:Xineq}
\end{equation}%
Next consider the $n=1$ case of Eq.~(\ref{eq:betaZ(n)}), and set it to be
smaller than $\beta _{Z}$:%
\begin{equation}
\beta _{Z}^{(1)}/\beta _{Z}\leq (\delta \beta _{X})\left[ \frac{2}{\pi }(1+%
\frac{1}{b})+\frac{a}{b}\right] +\frac{\delta }{\tau _{0}}<1.
\end{equation}%
Both quantities $\delta \beta _{X},\delta /\tau _{0}$ are $\ll 1$ by our
previous assumptions, so this inequality is automatically satisfied.
Finally, consider the $n=1$ case of Eq.~(\ref{eq:betaY(n)}), and set it to
be smaller than $\beta _{Y}$:%
\begin{equation}
\beta _{Y}^{(1)}/\beta _{Y}\leq (\tau _{0}-\delta )\beta +(\tau _{0}-2\delta
)\frac{b}{a}\beta _{X}+\frac{1}{\pi }\frac{\delta }{\tau _{0}}\frac{b}{a}<1.
\label{eq:Yineq}
\end{equation}

We can simplify these results somewhat, as follows. We can replace $\beta
_{X}$ by $J=\max (\left\Vert B_{X}\right\Vert ,\left\Vert B_{Y}\right\Vert
,\left\Vert B_{Z}\right\Vert )<\beta $ [Eq.~(\ref{eq:Jb})], and assume for
simplicity $a=b=1$. We can then replace Ineqs.~(\ref{eq:Xineq}),(\ref%
{eq:Yineq}) by%
\begin{eqnarray}
2(\tau _{0}-\delta )\beta +(\frac{1}{2}+\frac{1}{\pi })\frac{\delta }{\tau
_{0}} &<&1 \\
(\tau _{0}-\delta )(\beta +J)-\delta J+\frac{1}{\pi }\frac{\delta }{\tau
_{0} } &<&1.
\end{eqnarray}%
We can safely replace $J$ by $\beta $ (since $J<\beta $) and drop $\delta J$
. This turns the second inequality into $2(\tau _{0}-\delta )\beta +\frac{1}{
\pi }\frac{\delta }{\tau _{0}}<1$, so it is subsumed by the first inequality.

\section{Concatenation of the CPMG pulse sequence}

\label{appC}

Consider an error Hamiltonian $H_{e}$ of the form
\begin{equation*}
H_{e}=B_{z}\otimes Z+B_{0}\otimes I_{S}.
\end{equation*}%
A simple dynamical decoupling sequence (also known as CPMG \cite{Carr:54,Meiboom:58}) can
decouple this error Hamiltonian:
\begin{equation*}
\mathtt{p}_{1}=X\mathtt{f}X\mathtt{f}
\end{equation*}%
where $\mathtt{f}$ denotes a free evolution interval of duration $\tau $.
The propagator corresponding to this sequence is given by:
\begin{equation*}
U=e^{-i\tau (-B_{z}\otimes Z+B_{0}\otimes I_{S})}e^{-i\tau (B_{z}\otimes
Z+B_{0}\otimes I_{S})}.
\end{equation*}%
A simple application of the Baker-Campbell-Hausdorf (BCH) \cite{Reinsch:00}
expansion shows that $U$ can be written as
\begin{equation*}
U=e^{-i2\tau (B_{0}\otimes I_{S}+F[B_{0},B_{Z}]\otimes Z)}
\end{equation*}%
where $F[B_{0},B_{Z}]$ is a Hermitian operator in the Lie sub-algebra
generated by $B_{0}$ and $B_{Z}$. One can show that $\Vert
F[B_{0},B_{Z}]\Vert =O(\Vert B_{0}\Vert \Vert B_{Z}\Vert \tau )$ in the
limit of $\tau \rightarrow 0$. We can thus use the same sequence for
decoupling the undecoupled term $F[B_{0},B_{Z}]\otimes Z$ and construct 2nd
and 3rd order CDD sequences:
\begin{align}
\mathtt{p}_{2}& =\mathtt{f}X\mathtt{f}\mathtt{f}X\mathtt{f} \\
\mathtt{p}_{3}& =X\mathtt{f}X\mathtt{f}\mathtt{f}X\mathtt{f}X\mathtt{f}X%
\mathtt{f}\mathtt{f}X\mathtt{f}
\end{align}%
Notice that concatenated CPMG requires far fewer pulses than concatenated
universal DD. However, CPMG is not as robust with respect to systematic
errors in the pulses as concatenated universal DD.


\begin{thebibliography}{10}
\bibitem{CDD-PRA:note1}
This is not the only possibility; one may consider controlling the environment
  as well. See, e.g., A. Pechen and H. Rabitz, Phys. Rev. A \textbf{73}, 062102
  (2006).
\end{thebibliography}

\end{document}